\documentclass[epjST]{svjour}
\usepackage{graphicx,subcaption,capt-of}
\usepackage{placeins}
\usepackage{chemformula}

\begin{document}
\title{Edge-Detected 4DSTEM - effective low-dose diffraction data acquisition method for nanopowder samples in a SEM instrument}

\author{Nikita Denisov\inst{1,2}\thanks{\email{nikita.denisov@uantwerpen.be}} \and Andrey Orekhov\inst{1,2} \and Johan Verbeeck\inst{1,2} }
%
\institute{EMAT, University of Antwerp, Groenenborgerlaan 171, Antwerp, 2020, Belgium\and Nanolab center of excellence, University of Antwerp, Groenenborgerlaan 171, Antwerp, 2020, Belgium.}
\abstract{
    The appearance of direct electron detectors marked a new era for electron diffraction. Their high sensitivity and low noise opens the possibility to extend electron diffraction from transmission electron microscopes (TEM) to lower energies such as those found in commercial scanning electron microscopes (SEM).
    The lower acceleration voltage does however put constraints on the maximum sample thickness and it is a-priori unclear how useful such a diffraction setup could be. On the other hand, nanoparticles are increasingly appearing in consumer products and could form an attractive class of naturally thin samples to investigate with this setup.
    In this work we present such a diffraction setup and discuss methods to effectively collect and process diffraction data from dispersed crystalline nanoparticles in a commercial SEM instrument. We discuss ways to drastically reduce acquisition time while at the same time lowering beam damage and contamination issues as well as providing significant data reduction leading to fast processing and modest data storage needs. These approaches are also amenable to TEM and could be especially useful in the case of beam-sensitive objects.
    } 
\maketitle
\section{Introduction}
    \label{intro}
    Electron diffraction has a long history in providing crystallographic data from materials, relying on the short wavelength that is easily obtained by accelerating an electron beam. Even at acceleration voltage as low as 1~kV we already obtain a matter wavelength of 38~pm which is more than adequate to reveal the interatomic distances in any material and comparable to common wavelengths used in X-ray diffraction setups. 
    Electrons have the benefit over X-rays in providing more information per incoming particle while at the same time resulting in less damage per information for a given volume of material \cite{henderson_potential_1995,gruene_rapid_2018}.
    
    The higher interaction strength of electrons compared to X-rays also has drawbacks as the penetration depth of electrons is considerably less and multiple scattering complicates the interpretation of the diffraction patterns considerably \cite{vincent_double_1994,kolb_automated_2019}. This drawback turns into an advantage when particles become smaller and smaller as we currently experience with the growing application of nanoparticles in all aspects of life, but as is also commonly the case for all proteins and virus particles that are inherently small.

    For these reasons, electron diffraction in TEM is a widely used technique that is growing in success as the amount of structure solutions obtained with electron diffraction steadily increases in recent years \cite{klar_accurate_2023}. Electron diffraction did not only occur in TEM, but dedicated electron diffractometers were built and used in the past \cite{pinsker_electron_1953,pinsker_electron_1959} and such instruments seem to be reappearing now \cite{sieger_polymorph_2023,ito_structure_2021,mckenzie_establishing_2023} .

    In terms of technology, several breakthroughs came when moving away from photographic plates, to imaging plates to CCD camera's. In recent years direct electron detectors have emerged as nearly perfect detector for electron diffraction, showing near ideal quantum efficiency and near infinite dynamic range \cite{clough_direct_2014,levin_direct_2021,mcmullan_chapter_2016,faruqi_evaluation_2003,nord_fast_2020,tate_high_2016}.
Transmission electron microscopes were the natural instruments to apply this new detector technology but soon also experimental setups based on scanning electron microscopes started to appear with various levels of detector technology \cite{bogner_history_2007,caplins_transmission_2019,sun_progress_2018,orekhov_wide_2020,slouf_powder_2021,muller_exploring_2024,schweizer_low_2020}. The idea behind using an SEM instrument as a TEM diffraction tool hinges on the fact that to reach the far field plane which holds the diffraction pattern, a set of projector lenses as available in any TEM is not strictly necessary and simple free space propagation works just as well. Given the fact that a lot more space around the sample is available in an SEM, finding this propagation length is straightforward and propagating in free space comes with the advantage of avoiding geometric distortions that are typically caused in TEM by the projector lenses and need to be corrected to allow for quantitative data interpretation. Note that free space propagation is also what is used in X-ray diffraction albeit at typically higher scattering angles.
Added benefits of using an SEM instrument are the ease with which external stimuli can be integrated around the sample area, relative lower cost and lower barrier of entry as compared to TEM. Typically also the field of view can be substantially larger, allowing for far better statistics and automated routines as compared to TEM. This larger field of view and the ability to gather much more data, however comes with the challenge of high acquisition times.
In this paper, we demonstrate such a SEM diffraction setup and show ways to more efficient data collection from large numbers of particles and automatic information extraction from the acquired data.

\section{Materials and Methods}
\subsection{Experimental setup and samples}
    \begin{figure}[h]
        \centering
        \includegraphics[width=.4\textwidth]{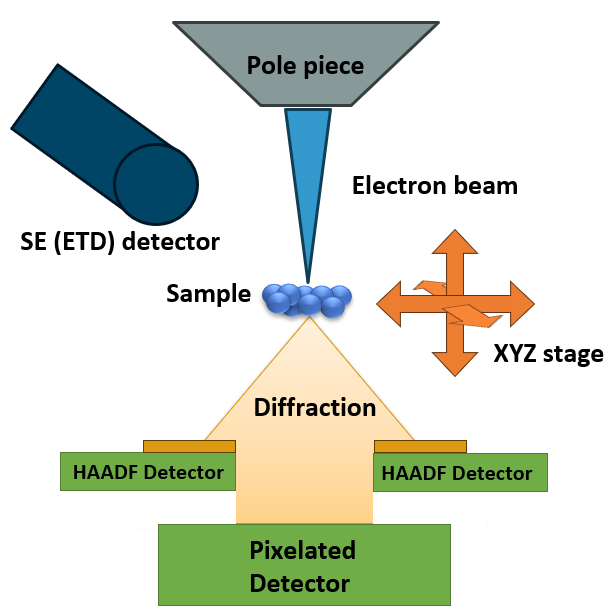}
        \caption{Experimental set up for transmission (diffraction) studies in SEM}
        \label{fig:set_up}
    \end{figure}

    The setup consists of a Tescan Mira FEG SEM outfitted with several modifications. Modifications include: Quantum Detectors scan engine for external control of the scan coils and reading the detectors' signals, custom-made open-source ADF detector \cite{vlasov_low-cost_2023},
    piezo XYZ stage assembly comprised of Xeryon piezo motors with encoder resolution of 312~nm, custom-made TEM sample grid holder and connector to stage, Advacam Advapix hybrid pixel detector with Timepix 1 300~$\mu$m Si chip with 256x256 pixels of 55~$\mu$m pitch and several custom-built feedthroughs to power and communicate with electronics inside of the microscope chamber.\par
    Software for experiment and setup control is made with the Python programming language using the open API offered by the SEM, scan engine, piezomotor controller and pixelated detector.\par
    A simplified schematic of the setup is presented in Fig.\ref{fig:set_up}. One of the advantages in such arrangement is the absence of a projection system in between the sample and the pixelated detector. Unlike TEM there is enough room in the microscope chamber of an SEM to freely change position of the sample and/or detector to magnify the diffraction pattern sufficiently. Minimal and maximal camera length reachable with our setup was from 10 to 90~mm with reciprocal space coverage from 1.1 to 2.26~$\text{\AA}^{-1}$ respectively. The limited 256x256 pixels of the detector resulted in a maximum angular resolution of 0.0043 to 0.0088~$\text{\AA}^{-1}$.
    \par
    Additionally we characterized the direct electron detector performance and established that its MTF and DQE are suitable for the low acceleration voltage range (15-30~keV) \cite{denisov_characterization_2023}.\par
    Test objects investigated in this work are \ch{SrTiO_3} lamella prepared with FIB, Lithium Iron Phosphate (\ch{LiFePO_4}) in the form of a nanopowder and single-crystal Silicon that was manually crushed to powder.\par
    Sample thickness studies were performed with a Titan (Thermo Fisher) transmission electron microscope outfitted with EELS detector.
    
\subsection{Edge Detected 4DSTEM method}
\label{sec:ed4dstem}
    \par
    A key challenge in low-voltage electron diffraction is the limited penetration depth in comparison to sample thickness.
    
\begin{figure}[h]
        \centering
        \includegraphics[width=.7\textwidth]{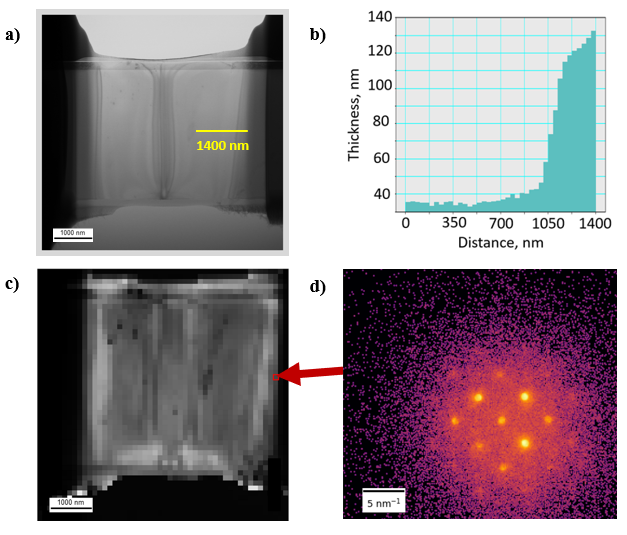}
        \caption{Estimation of transmission thickness limit at 30~keV accelerating voltage on STO lamella: a) TEM overview image with line profile for EELS thickness measurement, b) TEM EELS thickness profile of STO lamella, c) 4DSTEM virtual bright field image of STO lamella taken at 30~keV in SEM, d) Diffraction example at 120~nm thickness position of EELS line profile showing that even at 120~nm a reasonable quality diffraction can be obtained.}
        \label{fig: Thickness}
    \end{figure}

\begin{figure}[h]
        \centering
        \includegraphics[width=.7\textwidth]{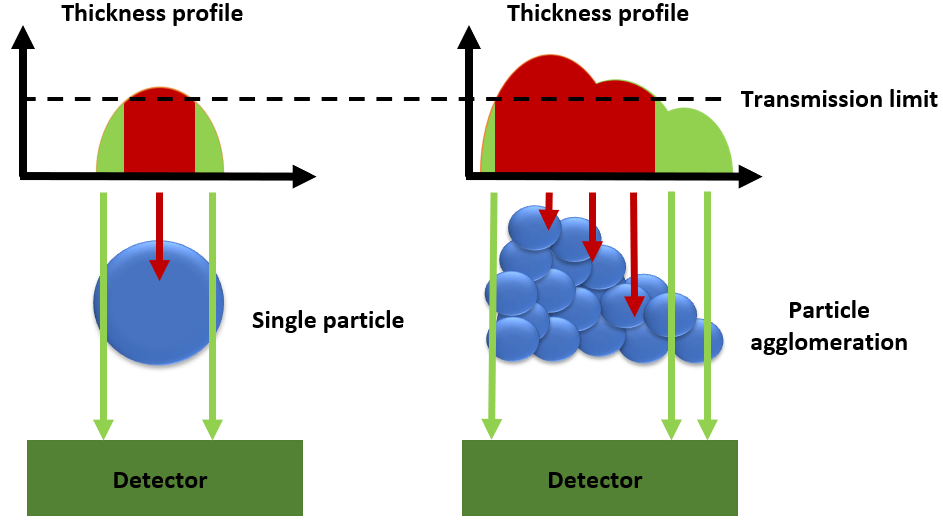}
        \caption{Sketch of 2 situations where the particle or particles agglomerate is too thick to result in transmission diffraction patterns. Note that in both cases, the perimeter around the (agglomorate of) particle(s) still provides areas where the thickness is much lower resulting in good quality diffraction patterns.}
        \label{fig: EDMethod}
    \end{figure}

    The inelastic mean free path is given as \cite{malis_eels_1988,zhang_eels_2011}
\begin{eqnarray}
\lambda=\frac{106 F E_0}{E_m ln(2\beta E0/E_m)}
\end{eqnarray}
With $E_m=7.6 Z^{0.36}$ and F the relativistic factor. This scales approximately as $E_0$ for nonrelativistic energies and results in a significant reduction of the penetration depth for e.g. 30~keV as compared to the typical range of energies used in TEM (80-300~keV). On the upside, the nonrelativistic interaction constant: 
\begin{eqnarray}
\sigma=\frac{\pi}{\lambda E_0}
\end{eqnarray}
Scales inversely with beam energy and provides more information per thickness which is a major advantage for investigating small particles, even more so when comparing to X-rays.
The downside is however that not all particles are naturally thin and would require more sample preparation steps as would be required in TEM. To avoid changing the structure of the particles, we however come up with an alternative solution that works very well in some cases.
In order to demonstrate this thickness issue we prepared a TEM lamela of \ch{SrTiO_3} in [001] crystallographic orientation with FIB and deliberately created a thickness ramp by thinning the middle part to 30~nm that we characterized in a TEM with EELS as shown in Fig.\ref{fig: Thickness}. We discovered that at 30~keV, we still get very acceptable diffraction patterns up to thicknesses of at least 120~nm in STO. This demonstrates that on the one hand, diffraction can deal with relatively thick samples but also that at some thickness the sample ceases to be transparent and we get no meaningful (empty) diffraction patterns.
For a typical powder sample, we could be in a situation where not all particles are electron transparent and especially with agglomerated particles this can become a real issue. Also for powder samples, the coverage of the TEM grid is never 100\%. In both cases, we would be wasting electrons by either recording patterns where no sample is present or where the sample is too thick to produce a diffraction pattern. This not only takes time while returning no information but it also adds to the total beam damage and might result in contamination build up in a less than ideal vacuum system.
To overcome these issues, we propose an alternative solution that we label as Edge Detected 4DSTEM (or ED4DSTEM for short). If we look at Fig.\ref{fig: EDMethod} we see two general nanoparticle sample cases - a single thick particle and an agglomeration of smaller particles. We see that in both cases, we might have issues getting meaningful diffraction patterns from the center of both arrangements, while on the perimeter we should get areas of adequate thickness.
The underlying assumption here is that we accept that the thick areas can't give meaningful information anyway but that the thinner parts are still representative of the whole sample. Depending on the type of samples, this may (homogeneous particles) or may not (e.g. core shell particles) be true. The strategy we propose now is to only visit those areas within a given thickness range based on a fast image acquisition that precedes the diffraction recording experiment.\par
    
    We follow the below procedure:
    \begin{enumerate}
    \item{Acquire an overview image of a region of interest in a "fast" manner - using a short pixel dwell time with either an SE or ADF detector.}
    \item{Apply a denoising filter on the overview image.}
    \item{Perform edge detection and create a scan position mask to control the scan engine.}
    \item{Apply a dilation filter to compensate for sample drift during the aquisition time and compensate for small beam positioning errors.}
    \item{Scan over the dilated scan mask position while acquiring high quality diffraction data with a pixelated detector at a longer dwell time.}
    \end{enumerate}\par
    
\begin{figure}[h]
        \centering
        \includegraphics[width=.6\textwidth]{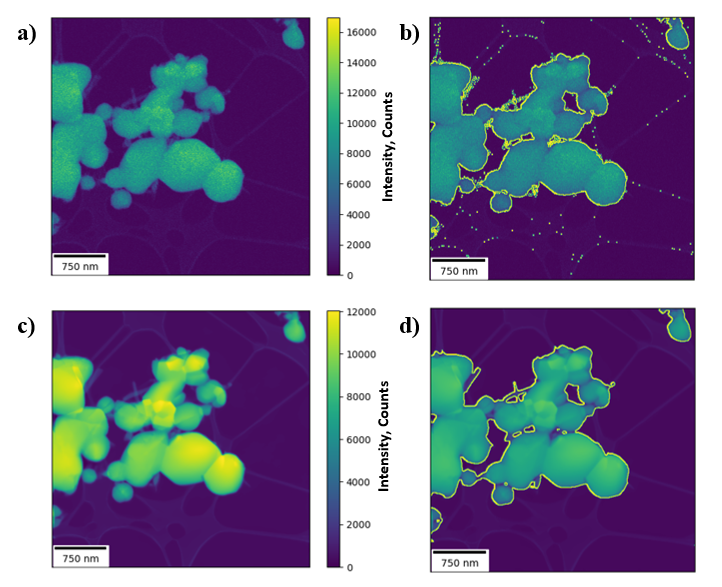}
        \caption{Effect of denoising on edge detection quality: a) overview image taken at 30~keV 50~pA 1~$\mu$s, b) scan mask (yellow) calculated via edge detection from image a), c) overview image denoised with \cite{lobato_deep_2024}, d) scan mask (yellow) calculated via edge detection from denoised image c).  }
        \label{fig: denoising}
    \end{figure}

Electron detection comes with inevitable counting noise, even in cases where all other sources of additive noise from the detector are kept to a minimum. As we aim to keep sample damage and contamination as limited as possible, we utilize a low beam current ($\approx$30-50~pA) together with low pixel dwell time ($1~\mu$s) for the fast overview scan. This leads to pronounced noise in our overview images. In Fig. \ref{fig: denoising}~a) we show an example of such overview image acquired with 1~$\mu$s dwell time 50~pA beam current and 30~keV acceleration voltage. In order to better decide which areas are worth visiting in the next step, we apply an image denoising algorithm (Fig. \ref{fig: denoising}~c). Nowadays a great number of denoising techniques are available \cite{fan_brief_2019}. We attempted several alternatives and found best results with an open source convolutional neural network developed in our group and described in \cite{lobato_deep_2024}. The result of this step is shown in Fig.\ref{fig: denoising}~c). 

We now compare the result of an edge detection algorithm on both the original overview image and the denoised version as shown in Fig.\ref{fig: denoising} b),d). It is clear from this example that without denoising the scan mask contains lots of artefacts, while this issue is strongly suppressed in the denoised version. Note that denoising here only helps to decide which regions to visit and no denoising is applied to our actual diffraction datasets.
    
The edge detection process involves using adaptive thresholding turning the denoised overview image in a black and white binarised image making use of the Otsu method \cite{otsu_threshold_1979}. Other simpler thresholding techniques (e.g. mean or median threshold \cite{mardia_spatial_1988}) were attempted as well but the adaptive thresholding holds the benefit of coping better with differences in image intensities e.g. when changing acquisition or instrument parameters.

    Next, edge detection is performed with a Canny detector \cite{canny_computational_1986} available in the openCV image processing package \cite{pulli_realtime_2012}. In order to improve robustness with respect to sample/microscope drift, we further dilate this mask homogeneously in all directions. The size of the dilation kernel is a trade-off between robustness and introducing areas which again are either too thin or too thick to result in meaningfull diffraction data. Performing a logical AND operation between dilated mask and the binarised thresholded image we obtain a scan mask that avoids scanning outside the particle.

\begin{figure}[h]
        \centering
        \includegraphics[width=.7\textwidth]{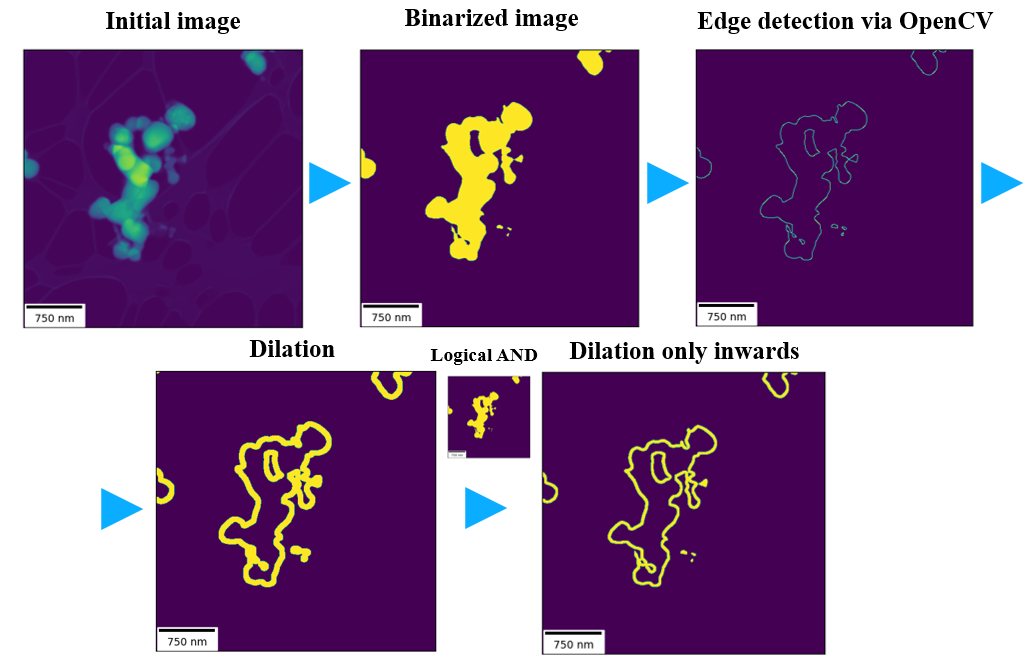}
        \caption{Illustration of the process steps to obtain the ED4DSTEM scan positions mask.}
        \label{fig: mask_creation}
    \end{figure}

    Finally the resulting scan mask is uploaded to a versatile scan engine. The entire scan mask creation process is illustrated in Fig. \ref{fig: mask_creation}.
    \par

\subsection{Diffraction data treatment}
The benefit of obtaining more information from smaller particles also comes with the challenge that these particles need to be supported. TEM grid support is typically an amorphous film. Looking at Fig.\ref{fig: Si_rings} a) we can barely distinguish any diffraction peaks in the Si mean diffraction pattern as they are drowned in the radially symmetric background originated from pronounced scattering from the amorphous support and investigated material. Indeed decrease in acceleration voltage leads to an increase in the inelastic scattering cross-section \cite{lobato_accurate_2014}.  However as we collect diffraction patterns in each probe position we can pre-process each pattern by finding the individual diffraction peaks (Fig.\ref{fig: Si_rings} d) thus avoiding accumulation of diffused diffraction background.

An alternative solution would be to reduce the thickness of the support e.g. by using graphene based sample grids \cite{pedrazo-tardajos_direct_2024,jain_exploring_2023} but even then it would be beneficial in terms of strong data reduction to perform peak finding on each individual diffraction pattern and storing only lists of positions and intensities. 
    Initially we performed peak finding with the Hyperspy Python package \cite{pena_hyperspyhyperspy_2024}. We found that due to a large amount of peak finding parameters which need to be varied from sample to sample, this solution was not well suited for automatic processing of significant quantities of data. In pursuit of a more robust solution we created a peak finding algorithm based on a modified version of PeakNet, a U-net neural network that was developed for X-ray diffraction\cite{wang_peaknet_2023}. We changed the neural network output processing to include diffraction peak positions irrespective of their diameter while initially it was finding peaks only for certain (small) diameters.
    A straightforward implementation without any optimisation (like parrallel or GPU computing) was already able to process one 256x256 diffraction frame in 4~ms. 
    \par
    
    The stored list of peak positions and intensities which results is a highly condensed dataset that holds all crystallographic information that was present in the experiment. This data can now be further used in existing algorithms for structure solution, orientation mapping, phase analysis etc. \cite{johnstone_pyxempyxem_2024,savitzky_py4dstem_2021,bucker_serial_2021}. In this work we display the results as a so-called virtual ring diffraction pattern (Fig.\ref{fig: Si_rings} b). Such ring pattern is obtained by plotting a scatter plot where each peak position is shown as a single spot with an intensity proportional to its diffracted intensity.
    When many spots are collected over many different grains or particles, we obtain a ring-like diffraction pattern that is similar to a typical powder diffraction pattern. Note that this way we get a much 'cleaner' diffraction pattern as nearly all amorphous and inelastic effects are removed in Fig.\ref{fig: Si_rings} b) as compared to Fig.\ref{fig: Si_rings} a) which shows a sum of all diffraction patterns recorded.
    Note that simple background subtraction of this summed pattern would not have been able to achieve the same result as the counting noise of the strong background signal has made it impossible to detect any of the weaker diffraction peaks.
    Another note - the virtual pattern does focus entirely on the crystalline content, which may be unwanted in some cases where eg. crystalline to amorphous ratio is important. In that case we could perform a hybrid solution where the summed pattern provides average information on the amorphous content while the virtual pattern still allows to identify the crystalline fraction. In e.g. powder XRD we would not be able to apply this trick as we are not able to get patterns from individual particles making it much harder to distinguish amorphous from cristalline content. On the other hand, we typically don't need a support in powder XRD either, but mixed amorhpous/cristalline samples are rather common.

\section{Results}
\subsection{Maximum sample thickness}
    To estimate the thickness limits imposed by the increased scattering at 30~keV we performed 4DSTEM and EELS analysis of an STO FIB lamella cut with [001] crystallographic orientation. The central part of the lamella was thinned down to 30~nm thus creating thickness ramp towards the sides.
    We find that the transmission maximum lies approximately in 120-130~nm thickness range (Fig.\ref{fig: Thickness}) for this sample at 30~keV. 
    
    As transmission depends among other parameters on density and crystallographic structure and orientation of materials, this result can not be generalised. We found however that for a wide range of different powders samples typical values for the transmission maximum were in the order of 90-140~nm for a 30~kV electron beam. This shows that for many true nanopowders, electron diffraction is feasible at this acceleration voltage and for larger particles our ED4DSTEM proposal would still find plenty of areas of this thickness range at the perimeter of even micron sized particles which are abundant. We note that support thickness should be kept as low as possible.

\begin{figure}[h]
        \centering
        \includegraphics[width=.6\textwidth]{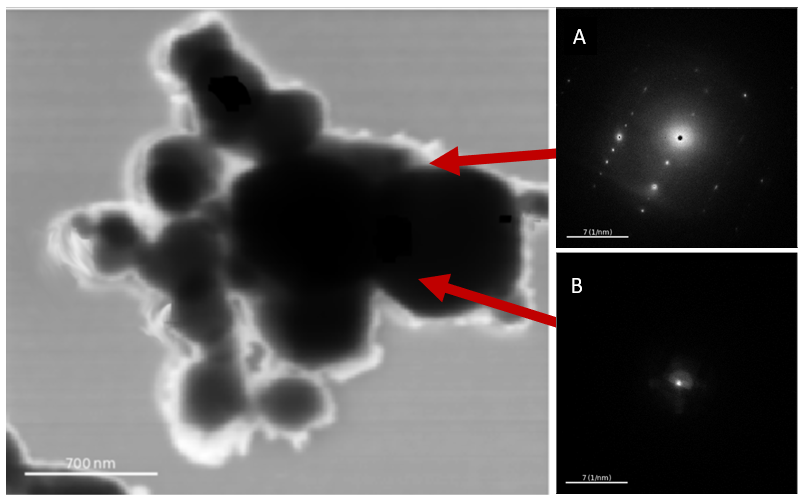}
        \caption{Lithium Iron Phosphate ($LiFePO_{4}$) 4DSTEM taken at HT = 30keV and BC = 50pA with diffraction examples: a) at the edge of agglomerate, b) at the middle part of agglomerate. }
        \label{fig: diff_comparison}
    \end{figure}

Fig.\ref{fig: diff_comparison} shows a typical agglomerated sample which demonstrates that indeed in the center of this ensemble we get uninformative diffraction patterns (b) as the sample is barely transparent. Nonetheless we still do observe diffraction patterns on the edges of the agglomerate (Fig.\ref{fig: diff_comparison} a) which shows these regions can still be used to analyse this sample even at acceleration voltages that are much lower than used in general TEM diffraction experiments \cite{jannis_event_2022,sari_analysis_2015}.\par
\subsection{Comparison of 4DSTEM and ED4DSTEM}

\begin{figure}[h]
        \centering
        \includegraphics[width=\textwidth]{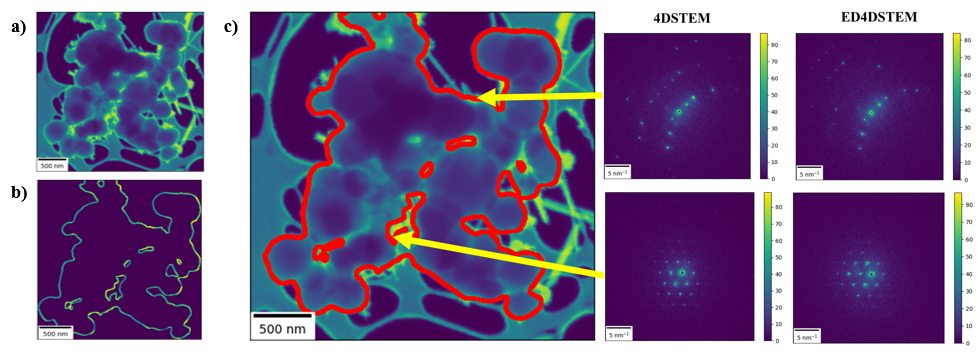}
        \caption{ a) 4DSTEM and b) ED4DSTEM images of Lithium Iron Phosphate taken at HT = 30~keV and BC = 50~pA, c) Merged 4DSTEM and ED4DSTEM (Red) images for LFP with comparison of diffraction taken by both methods at same sample position.}
        \label{fig: 4d_ED4d_diff_comparison}
    \end{figure}

    Precise estimation of the ED4DSTEM method efficiency is difficult due to different sample to sample parameters like  particle size and shape, its distribution and concentration, composition, etc. Thus we provide a qualitative comparison between ED4DSTEM and 4DSTEM methods based on one typical scan area of Lithium Iron Phosphate particle on carbon support. In Fig.\ref{fig: 4d_ED4d_diff_comparison} a) are shown a full scan with 4DSTEM and a ED4DSTEM scan (b) taken with same experimental parameters. In Fig.\ref{fig: 4d_ED4d_diff_comparison} c) we present an overlap of the 4DSTEM scan with the probe track of ED4DSTEM (red line) as well as examples of the diffraction patterns which demonstrate the same analyzed regions. It is seen that ED4DSTEM provides identical diffraction data and diffraction quality at the perimeter locations as normal 4DSTEM while being far more efficient in terms of dose, contamination, time for experiment and resources by skipping the parts that don't add useful diffraction data.\par  

    The method performance comparison of the full scan area (4DSTEM) and scan only 1- and 3-pixel perimeter thickness is provided in Tab.\ref{tab:comparison_numbers}. In the ideal case 1 pixel perimeter should be sufficient however as due to complex particle shape of weak signal from the thin particle edges or the microscope instability/drift we considered a case of 3 pixels broadening of the perimeter thickness. As we can see ED4DSTEM with 1- and 3-pixels perimeter is $\approx64$ and $\approx17$ times less points to visit compared to a conventional 4DSTEM. Similar results can be observed for the total time to complete experiment and the storage space - $\approx45$ and $\approx12$ times less respectively. Estimations of the total electron dose applied to the sample during experiments show a decrease by $\approx31.8$ and $\approx8.4$ times. This gain in electron dose efficiency depends on the ratio of total area to perimeter area of the particle or agglomerate. We expect it to scale with the size of investigated object - the larger the object - the larger the dose efficiency of ED4DSTEM compared to 4DSTEM.\par

    \begin{table}[h]
    \caption{4DSTEM and ED4DSTEM performance comparison with same experimental conditions for LFP agglomerate.}
    \label{tab:comparison_numbers}    
    \begin{tabular}{c|cccc}
    \hline\noalign{\smallskip}
    & Amount of frames & Total electron dose & Time,Sec & PC storage, MB\\
    \noalign{\smallskip}\hline\noalign{\smallskip}
    4DSTEM & 154 368 & 241.97~*$10^8$ & 232 & 3756\\
    ED4DSTEM 1px thick & 2 429 & 7.59~*$10^8$ & 3.7 & 82.8 \\
    ED4DSTEM 3px thick & 9 213 & 28.96~*$10^8$ & 14 & 314 \\
    \noalign{\smallskip}\hline
    \end{tabular}
    \end{table}
\subsection{ED4DSTEM method diffraction data evaluation}

    \begin{figure}[h]
        \centering
        \includegraphics[width=.8\textwidth]{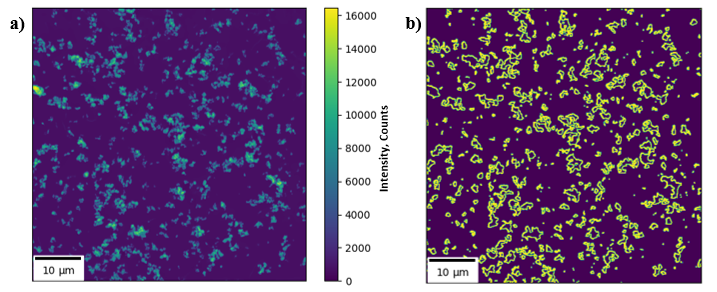}
        \caption{Example of large area (FOV 60~$\mu m^2$) ED4DSTEM experiment on Si powder: a) ETD overview image, b) Detected scan mask for ED4DSTEM experiment}
        \label{fig: 4d_ED4d_large_area}
    \end{figure}
    
    \begin{figure}[h]
        \centering
        \includegraphics[width=.8\textwidth]{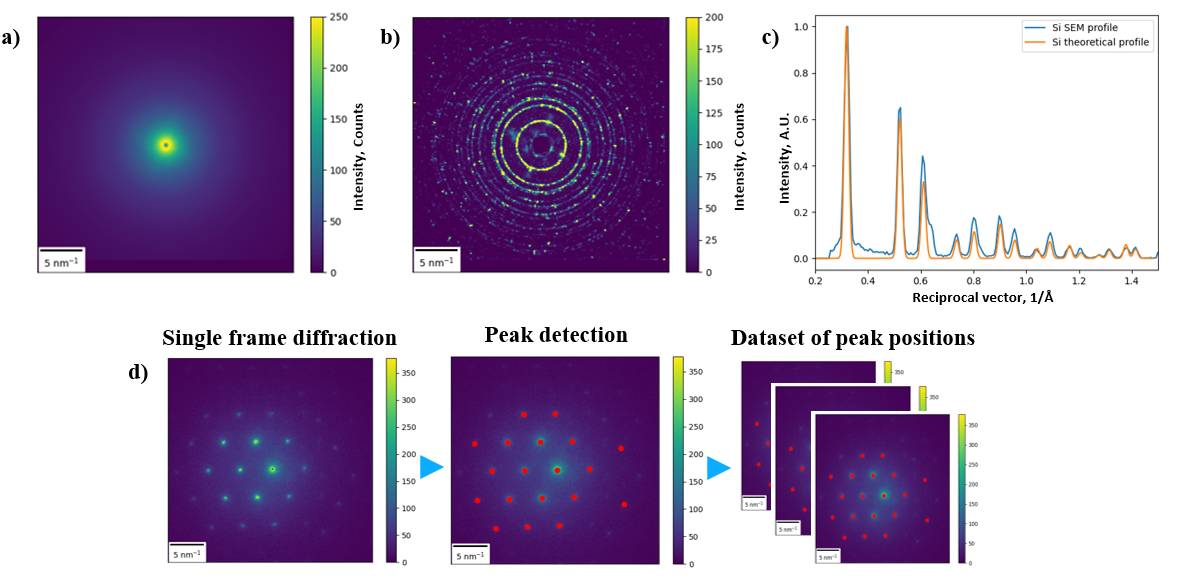}
        \caption{ED4DSTEM Diffraction data processing and data evaluation with theoretical diffraction profile on 138.4~k frames dataset acquired from Si powder at 30~keV 50~pA 1~ms exposure:  a) Si mean diffraction pattern without pre-processing, b) Si virtual mean pattern after peakfinding in each individual experimental frame,
        c) Radial average profile of ED4DSTEM Si ring pattern (Blue) in comparison with theoretical Si diffraction profile (Orange), d) Pipeline for creation of dataset of peak positions and/or peak intensities.}
        \label{fig: Si_rings}
    \end{figure}
    
    Now as we have established the process workflow, we apply it to a test case sample of nanoscristalline Si that was obtained by crushing single-crystal Si wafer in a mortar. In Fig.\ref{fig: 4d_ED4d_large_area} we present an example of one of the many regions from the large field of view ($\approx60~\mu m^2$) that was investigated along different areas of a TEM grid.\par
    We acquired in total 138.400 diffraction frames at acceleration voltage 30~keV, beam current 50~pA and exposure time of 1~ms per diffraction frame. Experiment took 180~sec to complete excluding time for stage repositioning to new regions. The resulting virtual ring diffraction pattern and radial average diffraction profile are presented in Fig.\ref{fig: Si_rings}. While comparing the experimental Si diffraction profile obtained in SEM with the theoretical Si profile calculated with CrystalDiffract package assuming the known structure for silicon from CIF library, we can see an almost perfect match of peak positions and intensity ratios. 

\section{Discussion}
    Results of comparison ED4DSTEM with 4DSTEM have shown how a simple modification of 4DSTEM dramatically increases effectiveness of diffraction studies both in terms of experimental requirements and electron dose applied to the sample.\par
    While using ED4DSTEM we have to keep in mind that during the experiment we visit the outer parts of the sample and we assume edges of our sample are representative for the whole. It is known that structure on edges for nanoparticles can be different from the bulk structure of a sample \cite{yokoyama_chapter_2008}. Moreover nanoparticles could be encapsulated or covered with various coatings that will prevent accurate sample analysis unless the goal is to investigate only these coatings. However, in cases where the assumption is acceptable, ED4DSTEM holds the attractive property of providing efficient statistical and reproducible diffraction experiments on nanoparticles and nanopowders in the affordable environment of a commercial SEM instrument.\par

It is interesting to compare this setup to a typical X-ray powder diffraction instrument. If we look at the fundamental limits of diffraction at nanoparticles we can write the limit to angular resolution as:
    \begin{eqnarray}
    \Delta \theta \propto \frac{\lambda}{d}
    \end{eqnarray}
    With $\lambda$ the wavelength and $d$ the minimum of either particle diameter or probe size.
    This angular broadening has to be compared to a given Bragg angle:
    \begin{eqnarray}
    \theta \propto \frac{\lambda}{a}
    \end{eqnarray}
    With a the interatomic distance of interest. This makes the relative broadening scale as:
    \begin{eqnarray}
    \frac{\Delta \theta}{\theta} \propto \frac{a }{d}
    \end{eqnarray}
    Independent of the wavelength and valid for both X-ray and electron diffraction if we can create a parallel enough incoming beam with opening half angle $\alpha$:
    \begin{eqnarray}
    \alpha < \frac{d}{\lambda}
    \end{eqnarray}
    With $d$ the diameter of the particle of interest. If we make the convergence angle larger, we can gain spatial resolution at the expense of angular resolution which is typically hard or impossible to do for X-rays for particles smaller than a few 100~nm.
    The only real difference between XRD and ED is then the interaction strength which is highly favorable for ED in case of small particles \cite{henderson_potential_1995}.
    In our experimental case, we suffer from a less than ideal electron detector that limits the angular resolution to $\Delta \theta \approx 0.02 \text{\AA}^{-1}$ due to the relatively low amount of pixels we have. If we assume a similar high pixel count detector as is used in a common XRD instrument, we should fundamentally get similar angular resolution.
    Note how the experiment already shows a fairly acceptable XRD-like powder pattern obtained from an estimated total volume of less than $0.01 \mu m^3$ which would be highly challenging for an X-ray diffractometer.
    
    Note that the proposed ED4DSTEM method is not limited to SEM only. TEM users can benefit from it as well for specific cases of large particles or agglomerations that cannot be fully transmitted, the largest difference being a thickness scaling with acceleration voltage.

    \par
    In order to get best results we used and recommend using continuous ultra-thin (3~nm thickness) TEM sample grids or similar means of suspending the powder sample. Continuous grids provide a homogeneous background and reliable contrast with the sample particles in the overview image which helps to determine the optimal experimental scan positions while minimum thickness of the support material limits presence of amorphous background in diffraction images.\par
    The ED4DSTEM scan mask due to the nature of edge detection algorithms can be created from any type of overview image that has sharp contrast between background and investigated object. In the current work we presented examples of overview images obtained with SEM secondary electron detector (Fig.\ref{fig: denoising} a,c) and HAADF detector (Fig.\ref{fig: 4d_ED4d_large_area} a) and their respective ED4DSTEM scan masks (Fig.\ref{fig: denoising} b,d and Fig.\ref{fig: 4d_ED4d_large_area} b). Another possible way of creating the ED4DSTEM scan mask apart from edge detection is to retrieve the scan mask directly from the HAADF detector signal which by design already indicates areas where transmission is possible. However the HAADF signal arises from highly scattered electrons and does not always represent positions in which direct diffraction with pronounced diffraction reflexes is present. Moreover the ADF signal is dependent on sample atomic number as well as collection angle (distance from sample to ADF detector) \cite{ishikawa_quantitative_2014,krivanek_gentle_2010} which means it might be an inconsistent reference for series of different samples while edge detection is not bound to these issues and do not require installation of specialized detector into SEM.\par
    
    The proposed data analysis pipeline, switching from the common image based post-processing (background removal from mean diffraction pattern) to a pre-processing by peak finding on individual frames and then forming virtual mean ring diffraction patterns significantly improves signal quality. Peak finding on individual frames can be performed "on the fly" during data acquisition. This holds the benefit of providing real time data that is already significantly reduced in size, which makes all subsequent steps far lighter and faster to implement. We stress here that this is not only a data reduction benefit, but because of the Poisson counting statistics having this position correlated data allows for a far better discrimination between crystalline and amorphous background which can not be obtained when averaging all data together as would be the case in a large probe electron diffraction or XRD experiment.

    Further progress on the detector side could be achieved by changing to an event-based detector like Timepix 3. Event-based acquisition further reduces the amount of data that needs to be stored as well as shortens the acquisition time as the dead-time of camera in between of acquisitions is absent for event-based detectors \cite{jannis_event_2022,poikela_timepix3_2014}.One drawback of such cameras could be their limited event rate that will likely force us to reduce the beam current or add a beamstopper.
    
\section{Conclusions}
    In this work we proposed Edge Detected 4DSTEM - a modified 4DSTEM method for use with low acceleration voltage conditions. We have shown that for low acceleration voltage it is possible to collect direct diffraction only from the edges of the sample due to presence of sufficiently thin regions in those areas.\par
    We performed a comparison between ED4DSTEM and a more conventional 4DSTEM experiment and showed significant efficiency improvement and decrease in electron dose applied to the sample by visiting specific regions of the sample while obtaining identical data quality.\par
    We presented an approach to ED4DSTEM data treatment - peak finding on individual frames with later formation of virtual mean diffraction images. Such approach allows to avoid background subtraction, retrieve the weakest diffraction reflexes and to identify small portions of crystalline phases in amorphous-crystal mixtures.\par
    Finally we demonstrated the ED4DSTEM diffraction results on a \ch{Si} powder sample by comparing the ring diffraction profile retrieved in the ED4DSTEM experiment with simulated \ch{Si} diffraction profile. The ring diffraction profile matched well both in peak positions and peak ratios showing the potential of the ED4DSTEM.\par
    We believe ED4DSTEM has great potential for use in large scale nanopowder and nanoparticle statistical studies. In combination with other techniques, for example SerialED, shape/size analysis, phase/orientation mapping, ED4DSTEM can open new horizons for automatized high throughput nanopowder analysis in the research and industrial fields where there is need for fast and reproducible structural experiments conducted on instruments that can be deployed widely.
    
\section*{Acknowledgments}
The authors acknowledge the financial support of the Research Foundation Flanders (FWO, Belgium) project number SBO S000121N.

\section*{Author contribution statement}
\textbf{Nikita Denisov:} Conceptualization, Methodology, Hardware, Software, Experimental, Original Draft. \textbf{Andrey Orekhov:} Conceptualization, Methodology, Hardware, Draft review and editing. \textbf{Johan Verbeeck:} Conceptualization, Methodology, Supervision, Draft review and editing. 

\bibliographystyle{unsrt}
\bibliography{Main_text_refs}

\end{document}